\documentstyle[11pt,paspconf,epsf]{article}
\begin{document}

\title{A Lithium Age for the Young Cluster IC~2391}

\author{David Barrado y Navascu\'es\altaffilmark{1}}
\affil{Max-Planck-Institut f\"ur Astronomie, Heidelberg,  Germany}

\author{John R. Stauffer and Brian M. Patten\altaffilmark{1}}
\affil{Harvard--Smithsonian Center for Astrophysics, Cambridge, USA}

\altaffiltext{1}{Visiting Astronomer, CTIO. 
 operated by AURA, Inc.\ under cooperative agreement with the National
Science Foundation}

\begin{abstract}
We have identified a large number
of possible very low mass members of the 
cluster IC~2391 based primarily on their location in an $I$ versus 
($R-I$)$_C$ CM diagram.  We have obtained
new photometry and low resolution ($\Delta \lambda = 2.7$ \AA\ )
spectroscopy of 19 of these objects (14.9 $\le$ $I_C$ $\le$ 17.5) in order 
to confirm cluster membership. We identify 15 of our targets as 
likely cluster members based on their  spectral types, 
radial velocity, EW(NaI8200\AA), and H$\alpha$ emission strengths. 
 One of these  stars has a definite lithium  detection and two other 
(fainter) stars have possible lithium detections. We find the lithium 
depletion boundary in IC~2391 is at $I_C$=16.2, which implies an age 
for IC~2391 of 53$\pm$5 Myr.  While this is considerably older than the 
age most commonly attributed for this cluster ($\sim$35 Myr), the
correction factor to the IC 2391 age is comparable to those recently 
derived for the Pleiades and $\alpha$ 
Per clusters and can be explained by new models for high mass
stars that incorporate a modest amount of convective core overshooting.
\end{abstract}

\section{Observations}

At a distance of $\sim$155 pc and with a canonical age of $\sim$35 Myr, 
IC~2391 is a ideal candidate to obtain an age based on the lithium depletion
boundary (LDB). This method is claimed to be 
less subject to possible systematic errors than
those based on isochrone fitting to the upper main sequence 
(Ventura et al. 1998a,b).
See Stauffer (1999, these proceedings) for more details about the 
LDB method.
We have carried out several observing campaigns to identify
new very low mass members, based primarily on their photometric
properties, and to confirm membership and the location 
of the LDB, via spectroscopy. 

{\sc Photometry:} (i) ``Big Throughput Camera'' at the 4m/CTIO (Jan 1998).
 We covered  2.5 sq. deg. in the $RI_C$ filters, reaching a limiting $I_C$=23,
 (ii) 0.9m/CTIO $+$ $VIZ$ filters
 (April 1998 and Jan 1999), (iii) Additional $JHK$ photometry from
 2MASS project (Adams 1998, priv. com).
 In total, we identified 126  candidate members
 located in the cluster loci in all CM and CC diagrams.

{\sc Spectroscopy:} CTIO 4m/RC spectrograph, Jan 1999 (6300-8820 \AA, 2.7 \AA\
resolution. We observed 19 IC~2391 candidate members, as well as several field
M dwarfs for comparison purposes.

Additional details about the photometry and the selection of members
can be found in Barrado y Navascu\'es et al. (1999).

\begin{figure}	
\vspace{-0.5cm}
\plotone{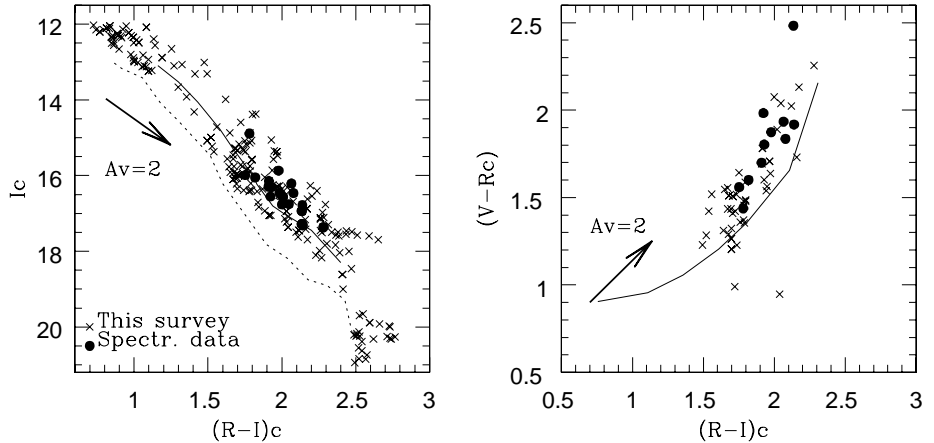}
\vspace{-7.0cm}
\caption{CM ({\bf a}) and CC ({\bf b}) diagrams for our sample of IC~2391
 candidate members.  Those objects observed spectroscopically appear
as solid circles.} 
\end{figure}

\section{The selection of members.}

The LDB age technique requires moderate S/N and spectral resolution.
Even for the closest and youngest clusters, the use of 4m class 
telescopes (or larger) and long exposures times (in the case of IC~2391, 
up to 4 hours) are necessary.
Therefore, it is vital to select the best membership candidates 
as targets for spectroscopic observations. This is achieved by using 
CM and CC diagrams (see Figure 1a,b).  With spectra in hand
(some examples are shown in Figure 2),  we used the
following criteria to determine the cluster membership status of our 
candidates:

\begin{itemize}
\item Radial velocities were compared with the cluster average of 15 km/s 
(Stauffer et al. 1997).
\item ($R-I$)$_C$ colors were derived by comparing the strength of several 
pseudo-continuum  bands (TiO, VO and CaH) with the values of field stars.
These colors were compared 
with the photometric values, assuming  E($R-I$)$_C$=0.01 for IC~2391.
The same method was used to derive spectral types.
\item  The gravity dependent NaI8200 \AA\ doublet (strong in dM, 
weaker in PMS M, and very  weak in M giant stars, Mart\'{\i}n et al. 1996)
was used to eliminate background giants and nearby field stars.
\item H$\alpha$ equivalent widths were compared with typical values 
for field stars and Pleiades members (see Figure 3a). This graph 
indicates that most of our objects are chromospherically very active
and, therefore, young. 
\end{itemize}

These criteria allowed us to reject two stars as members of the cluster,
and classify another 2 as possible non-members. The other 15 objects
are likely members of the IC~2391 cluster. The two faintest members 
are likely to be brown dwarfs, based on their photometry and theoretical 
models (D'Antona \& Mazzitelli 1997, Chabrier \& Baraffe 1997).

\begin{figure}	
\vspace{-0.5cm}
\plotfiddle{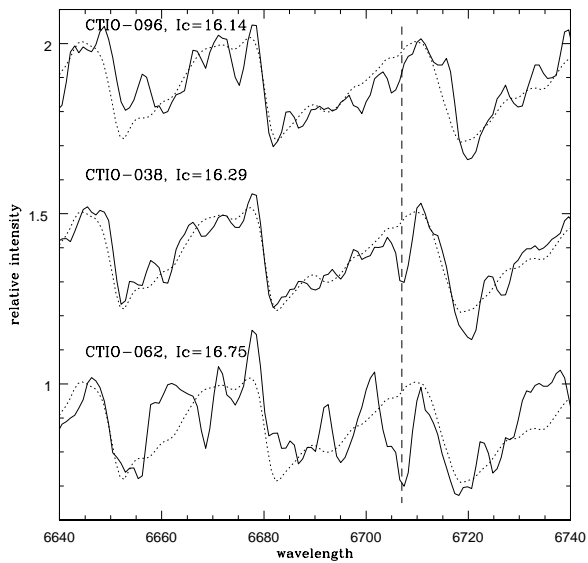}{7cm}{0}{40}{40}{-120}{-75}
\vspace{-0.0cm}
\caption{Spectra of IC2391 candidate members near the LDB.}
\end{figure}

\section{The lithium depletion boundary and the age of IC~2391.}

Lithium was detected in one of the bona-fide members and was possibly 
detected in two other bona-fide members (Figure 2).
Figure 3b shows a CM diagram for our candidates. We indicate those
having lithium as well as those members without it.  A ZAMS and two 
isochrones of 30 and 50 Myr are included for comparison 
(D'Antona \& Mazzitelli 1997). 
We estimate that the location of the  LDB is at $I_C$=16.2$\pm$0.15. 
Using a distance modulus of $m-M = 5.95 \pm 0.1$ and A$_I$=0.02 yields 
M($I$)$_{\rm LDB}$=10.23.  This implies a cluster age of 53$\pm$5 Myr, 
following Figure 3 of Stauffer et al. (1998).  Additional details on
this result can be 
found in Barrado y Navascu\'es, Stauffer, \& Patten (1999).
While this value of the cluster's age is significantly larger than previous 
estimates ($\sim$35 Myr), our result is consistent with recent age 
correction estimates of other young open clusters, such as the Pleiades
(125$\pm$8 Myr, Stauffer et al. 1998) and $\alpha$ Per 
(90$\pm$10 Myr, Stauffer et al. 1999), using the LDB method.
The LDB ages for these latter two clusters are consistent with age estimates 
using upper main sequence isochrone fitting using models which assume 
a modest  amount of convective core overshoot (Ventura et al. 1998a,b).
Since the ratio Age(standard)/Age(LDB) is almost constant for all these 
three clusters, it seems that 
the same  moderate amount of overshooting is needed (at least in the
mass range of the turn-off of these clusters). 
The new age scale for young open clusters based on the LDB ages 
would have important consequences in a large variety of topics, 
including the empirical and theoretical
correlations between lithium abundance, stellar activity, and rotation
rate as a function of age for low mass stars.

\begin{figure}	
\vspace{-0.5cm}
\plotone{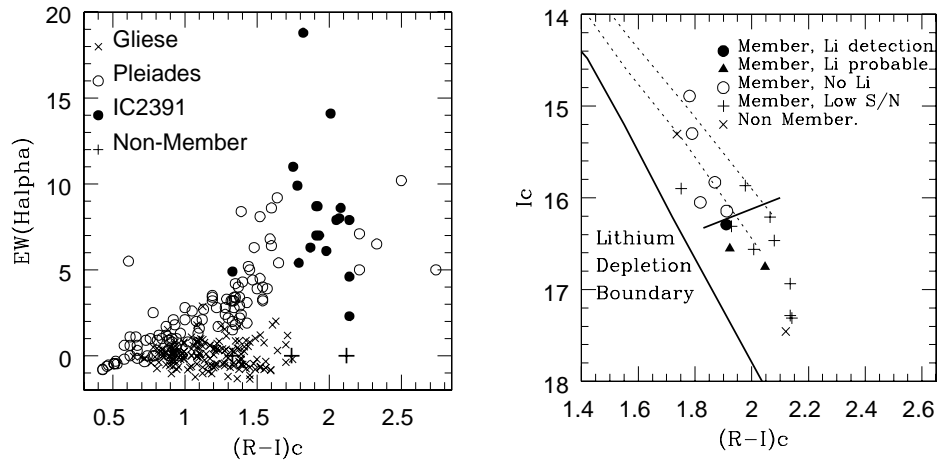}
\vspace{-7.0cm}
\caption{{\bf a} H$\alpha$ equivalent widths against the ($R-I$) color.
Bona-fide IC~2391 members are expected to show a relatively high level  
of chromospheric activity.  
{\bf b} CM diagram and the LDB for IC~2391.}
\end{figure}

\acknowledgments
DByN thanks the IAC (Spain) 
and the DFG (Germany) for their fellowship,
and the support by the European Union.

\end{document}